\documentclass[a4paper]{article}

\usepackage[english]{babel}
\usepackage[utf8x]{inputenc}
\usepackage[T1]{fontenc}

\usepackage[a4paper,top=3cm,bottom=2cm,left=3cm,right=3cm,marginparwidth=1.75cm]{geometry}

\usepackage{amsmath}
\usepackage{graphicx}
\usepackage{url}
\usepackage[colorinlistoftodos]{todonotes}
\usepackage[colorlinks=true, allcolors=blue]{hyperref}

\usepackage[small,labelfont=bf,up,textfont=it,up]{caption} 
\usepackage{booktabs} 
\usepackage{enumitem} 
\setlist[itemize]{noitemsep} 

\usepackage{siunitx}
\usepackage{xcolor}
\usepackage{authblk}

\title{\textbf{Concept validation of a high dynamic range point-diffraction interferometer for wavefront sensing in adaptive optics}}
\date{}
\author[1,2,*]{Nicol\'{a}s Dubost}
\author[1]{Nazim Ali Bharmal}
\author[1]{Marc Dubbeldam}
\author[1]{Richard M. Myers}

\affil[1]{Centre for Advanced Instrumentation, Department of Physics, Durham University, Durham, DH1 3LE, UK}

\affil[2]{SUPA, School of Physics and Astronomy, University of St Andrews, North Haugh, St Andrews, KY16 9SS, UK}

\affil[*]{\textit{nsdubost@gmail.com}}

\begin{document}
\maketitle

\begin{abstract}
The direct detection and imaging of exoplanets requires the use of high-contrast adaptive optics~(AO). In these systems quasi-static aberrations need to be highly corrected and calibrated. In order to achieve this, the pupil-modulated point-diffraction interferometer (m-PDI), was presented in an earlier paper. This present paper focuses on m-PDI concept validation through three experiments. First, the instrument's accuracy and dynamic range are characterised by measuring the spatial transfer function at all spatial frequencies and at different amplitudes. Then, using visible monochromatic light, an adaptive optics control loop is closed on the system's systematic bias to test for precision and completeness. In a central section of the pupil with 72\% of the total radius the residual error is 7.7\,nm~rms. Finally, the control loop is run using polychromatic light with a spectral FWHM of 77\,nm around the R-band. The control loop shows no drop in performance with respect to the monochromatic case, reaching a final Strehl ratio larger than 0.7.
\end{abstract}

\section{Introduction}
\label{sec:intro}
The future of optical astronomy poses a series of technical challenges, some of which are inherent to new telescope technologies. For example, all future Extremely Large Telescopes~(ELTs), some 10\,m class telescopes, such as the Gran Telescopio Canarias, and even the \textit{James Webb} Space Telescope have segmented primary mirrors. In order to perform as single monolithic mirror telescopes, active optics and wavefront sensing systems are required~\cite{yaitskova2003analytical,pizarro2002design,acton2012wavefront}. Other challenges are related to the new scientific cases these telescopes enable, when used in conjunction with technologies such as eXtreme Adaptive Optics~(XAO). One of the most prominent of these cases is the study of exo-planets. It has been estimated that, in systems such as the Gemini Planet Imager~\cite{fusco2006high}, non-common path aberrations~(NCPA) between the imaging arm and the wavefront sensor~(WFS) would have to be corrected to less than 10\,nm\,rms within the first 100 modes in order to achieve direct exo-planet detection. Furthermore, XAO systems need to achieve contrast levels of between $10^{-7}$ to $10^{-8}$ around 50\,mas and at rates of 2-4\,kHz~\cite{macintosh2006gemini,macintosh2008gemini}. In turn, for contrasts of $10^{-7}$ to $10^{-9}$, estimations indicate that the wavefront error~(WFE) would have to be lower than $\lambda /280$ and $\lambda /2800$ respectively, where $\lambda$ denotes the wavelength~\cite{stapelfeldt2005extrasolar}. This is equivalent to 6.0\,nm\,rms and 0.6\,nm\,rms in the H-band~($\lambda =1.65\,\mu m $).

Point-Diffraction Interferometers~(PDIs) such as the Zernike Wavefront Sensor~(ZWFS)~\cite{n2013calibration} and the self-referenced Mach-Zehnder~\cite{loupias2015development} have been proposed to tackle some of these challenges. Instead of performing indirect sensing like Shack-Hartmanns~(SHs), which reconstruct the wavefront from slope data, PDIs directly measure the phase without numerical reconstruction. Thanks to this, they have been found to be robust against optical vortices and other discrete phase errors~\cite{notaras2007demonstration} and are therefore well suited to measuring random piston and tip-tilt differences between telescope segments~\cite{yaitskova2005mach,janin2017fine}. Indeed, when phasing segments, ESO's Active Phasing Experiment showed them to be more robust than pyramid WFSs, SHs and curvature sensors with respect to seeing, to be able to perform both piston and tip-tilt corrections, and to achieve one of the lowest residual errors~\cite{gonte2009sky,surdej2010sky}.

Due to their nano-metric accuracy, PDIs have also been suggested to address the problem of accurately measuring and correcting quasi-static aberrations~\cite{bharmal2012interferometric,n2013calibration,wallace2010gemini}. For this purpose, some have been successfully tested on telescope-based systems, like ZELDA at the VLT-SPHERE instrument~\cite{n2016calibration}. ZELDA in particular is located at the focal plane, having similar advantages to focal plane WFSs such as the Self-Coherent Camera~\cite{mazoyer2013selfcoherentcamera}. Finally, it has been proposed that PDIs can help correct aberrations produced by atmospheric turbulence in XAO~\cite{love2005common,loupias2015development}, but testing is for the most part still limited to correction of static phase masks in the laboratory, under monochromatic conditions~\cite{evans2005extreme,loupias2016status}. This leaves XAO as one of the main challenges where PDIs are yet to produce promising laboratory and on-sky results.

The main drawback that so far prevents PDIs from becoming the main WFS in XAO systems is their small dynamic range, due to phase ambiguity. Without sufficient dynamic range, an AO system can not overcome the initial phase aberrations imposed by atmospheric turbulence, thus preventing closing the control loop. Methods such as the multiwavelength sweep~\cite{vigan2011sky,cheffot2020measuring} can be used to extend the dynamic range of PDIs up to several micrometers, but they do so by taking multiple acquisitions of the same phase aberration, each with different narrowly defined wavelengths. Both these requirements are difficult to achieve when measuring atmospheric turbulence with a natural guide star, making a single-frame large dynamic range WFS advantageous.


In our previous paper we presented a novel concept called the pupil-modulated PDI~(m-PDI)~\cite{dubost2018calibration}. In this paper we present experimental validations for this concept, embodied as the Complex Amplitude Wavefront Sensor~(CAWS). In Section~\ref{sec:principle} we give a brief overview of the instrument's working principle. Section~\ref{sec:sys} presents the design of the CAWS, as well as the experimental conditions provided by the CHOUGH~\cite{bharmal2018chough} bench.
Section~\ref{sec:range} introduces a new analytical formula to compute the instrument's dynamic range. The experiments performed on this bench were aimed at testing four different aspects of the CAWS: accuracy, dynamic range, completeness and chromatic bandwidth. In section~\ref{sec:transfer} we measure the spatial transfer function to test our model on the dynamic range and to quantify the instrument's accuracy. Then, Section~\ref{sec:loop} explores the CAWS' completeness by using it to compensate CHOUGH's systematic biases in closed-loop. Finally, Section~\ref{sec:poly} confirms that the CAWS can accurately estimate wavefront aberrations with broadband light, by closing the same control loop with polychromatic light.

\section{Working principle}
\label{sec:principle}
\begin{figure}[htb!]
  \centering\includegraphics[width=8cm]{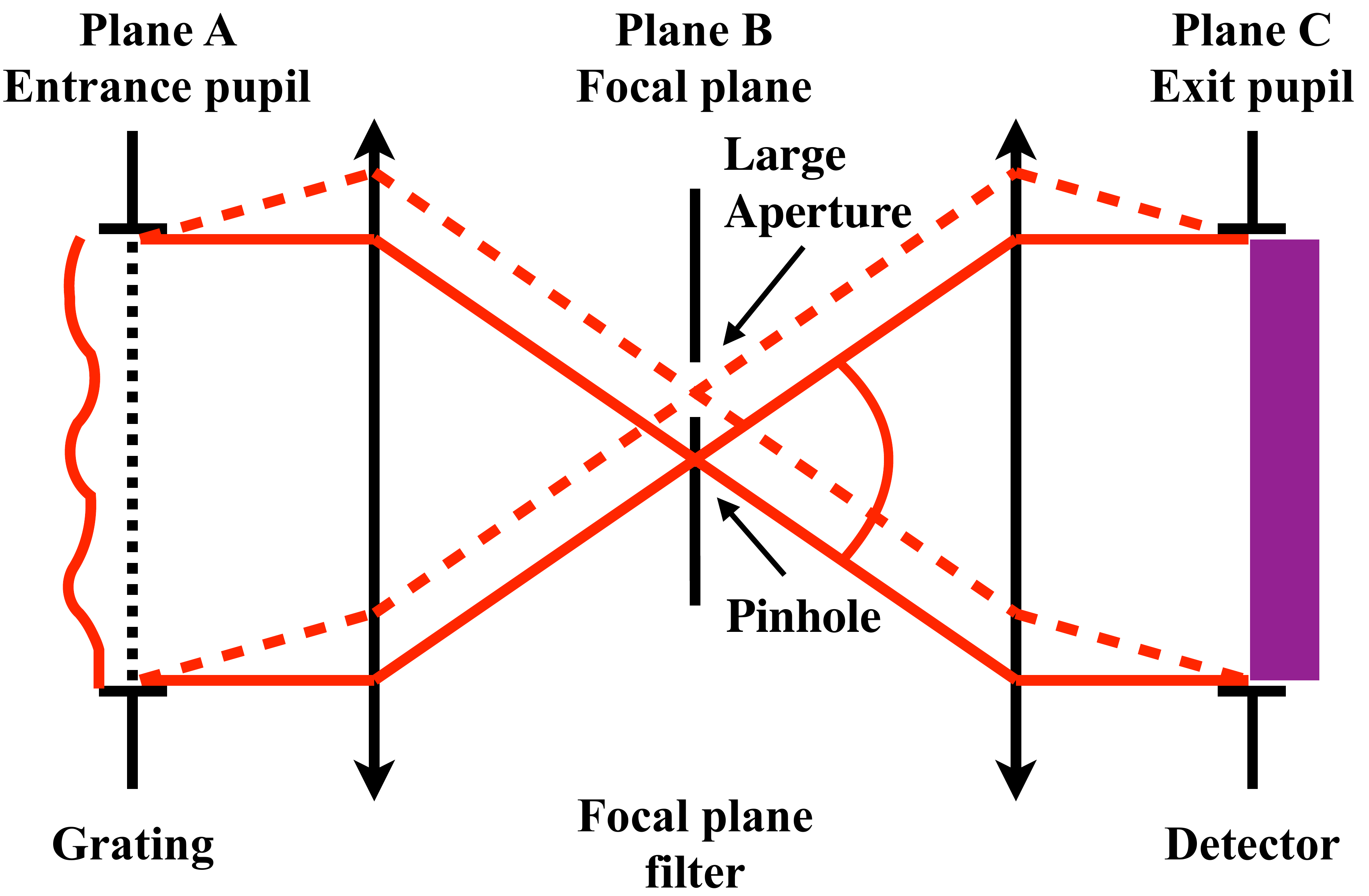}
  \caption{Layout of the pupil-modulated point-diffraction interferometer. The grating splits the beam into modes. Mode 0, shown in a solid red line, goes through a point-diffraction pinhole in the focal plane. Mode +1 goes through a larger aperture. The pupils and lenses are located as per a 4-f imaging relay, the basis for this design.
  }
  \label{fig:mpdi}
\end{figure}

We briefly recall the m-PDI principle~\cite{dubost2018calibration} on which the CAWS instrument relies to measure wavefront aberrations. For simplicity this introduction only considers the monochromatic case; the polychromatic case is studied in later sections. At the entrance pupil, the aberrated wavefront is split into at least two by a small-angle beam splitter, as presented in Figure~\ref{fig:mpdi}.
In the CAWS, this element is a diffraction grating which produces an infinite number of diffraction modes. These modes are focused onto a focal plane filter mask, which only allows mode~0~($M_0$) and mode~+1~($M_{+1}$) through. Furthermore, $M_0$ is filtered by a narrow pinhole in the mask in order to produce a flat reference beam. $M_{+1}$ goes through a larger aperture which filters all spatial frequencies to prevent cross-coupling, resulting in the test beam. Both beams are collimated and then interfered at the exit pupil on Plane~C. 

The resulting interferogram comprises fringes modulated by the incoming wavefront. This is due to the test beam arriving at a tilt and resulting in spatial carrier-frequency interference. The simplified expression for the intensity of the interferogram is
\begin{equation}\label{eq:simpleic}
  I_C(x,y) = P_0^2 + \frac{1}{\pi^2}P^2 + \frac{2P_0}{\pi}P \cos\left(\frac{2\pi}{T}x-\varphi_{LP}(x,y)\right),
\end{equation}
where $P$ and $\varphi _{LP}$ are respectively the amplitude and the low-pass filtered phase of the electric field $\Psi _0$ at the entrance pupil, $P_0$ is the average amplitude of the reference beam, and $T$ is the period of the interference fringes. Note that $T$ is proportional to the period of the grating and is the same for any wavelength. With respect to $P_0$ is in turn computed as
\begin{equation}\label{eq:p0}
P_0=P b_0 \sqrt{S},
\end{equation}
where $S$ is the Strehl ratio on the focal plane and $b_0$ is a number between 0 and 1 representing the fraction of the total amplitude, determined by the geometry of the pinhole and by the small-angle beamsplitter. In order to retrieve the phase $\varphi _{LP}$, the interferogram's intensity $I_C$ can be demodulated by performing a Fourier transform and retrieving a sideband. Performing a further Fourier transform on this sideband yields an estimation of the electric field, from which the phase $\varphi _{LP}$ can be computed~\cite{takeda1982fourier}. It is worth noticing that although the amplitude $P(x,y)$ can also be extracted from the electric field, this publication focuses exclusively on phase estimation.

\section{System description}
\label{sec:sys}
\begin{figure}[htb!]
\centering\includegraphics[width=13.2cm]{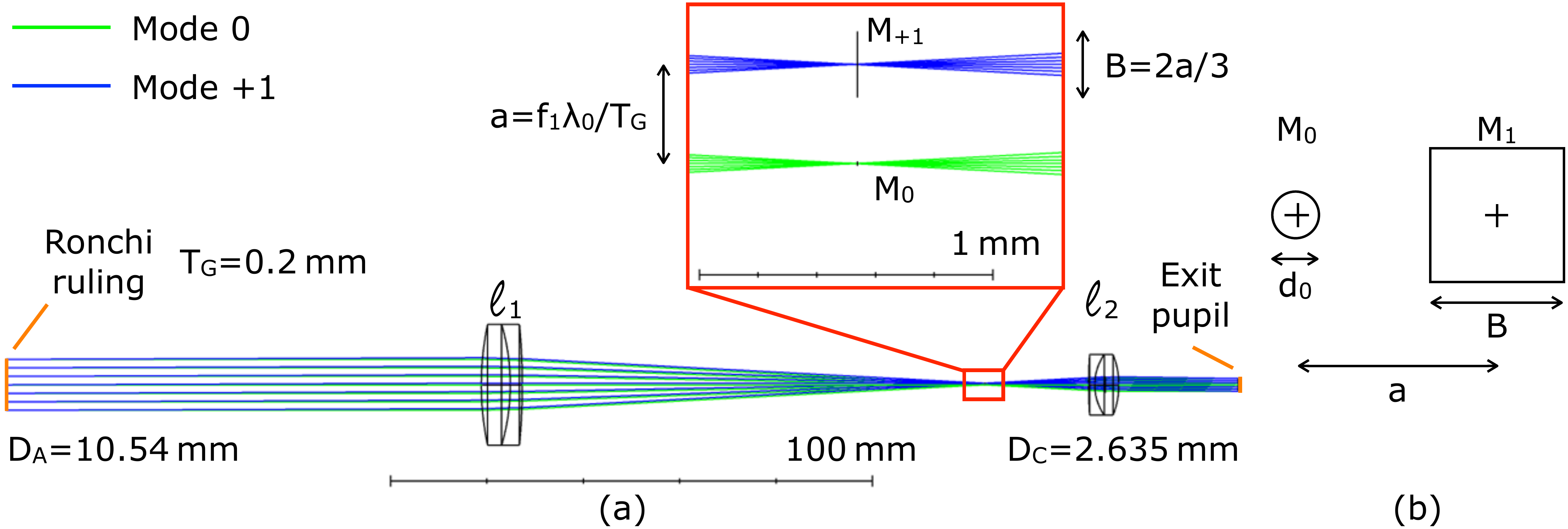}
\caption{CAWS' (a)~optical layout and (b)~diagram of the pinhole and aperture on the focal plane filter mask. In the red vignette, vertical black lines over $M_0$ and $M_{+1}$ represent the pinhole and aperture respectively. $T_G$ is the grating's period and $f_1$ is the focal length of lens $\ell_1$. The optical layout does not show more diffraction modes other than $M_0$ and $M_1$. The pinhole and aperture diagram in (b) is not to scale.}
\label{fig:layout}
\end{figure}

Figure~\ref{fig:layout} shows the optical layout of the CAWS.
The small-angle beamsplitter is a Ronchi diffraction grating that fits $52.7$ line-pairs in the pupil of diameter $D_A=10.54\,\mathrm{mm}$, as provided by CHOUGH. This gives it a spatial sampling of $17.5\,\rm{cycles/pupil}$. To put this into context, this is approximately the resolution of a 34-by-34~SH, which on a $4.2\,\mathrm{m}$ telescope such as the WHT would have $12.4\,\rm{cm}$ subapertures. The camera at the exit pupil has $7.4\,\rm{\mu m}$ pixels, resulting in an oversampled $6.8$\,pixel/line-pair sampling.

In the focal plane filter mask, $M_0$ is filtered by the pinhole resulting in the reference beam of the interferometer, and $M_{+1}$ goes through the larger sideband aperture which allows higher spatial frequency information through, resulting in the test beam. The diameter of the pinhole $d_0$ is $\SI{16}{\micro\meter}$, or $2.5 f_1 \lambda_0 /D_A$ with the pupil size delivered by CHOUGH and for a central wavelength $\lambda_0=675\,\mathrm{nm}$. With this pinhole size, $b_0=0.42$. From combining Eq.~\ref{eq:simpleic} and~\ref{eq:p0}, when $S \simeq 1$ the visibility of interference fringes is
\begin{equation}
V=\frac{2 b_0}{\pi b_0^2 + 1/\pi}=0.96.
\end{equation}

This first implementation of the instrument has a low throughput. From Eq.~\ref{eq:simpleic} and~\ref{eq:p0}, the average intensity of the output is
\begin{equation}\label{eq:meanic}
  \widehat{I}_C = \left(b_0^2 + \frac{1}{\pi^2} \right)P^2,
\end{equation}
which for $b_0=0.42$ is $\widehat{I}_C \simeq 0.28 P^2$, resulting in a throughput $\eta \simeq 0.28$. Note that half the light is lost on the Ronchi ruling, meaning the throughput of the focal plane mask is twice this value, or 0.56. This value also considers light lost to higher diffraction modes. Future iterations could more than double the throughput by implementing special transmission gratings.

From our previous paper, the side of aperture $M_{+1}$ is $B=2a/3$ where $a=f_1\lambda_0/T_G$, resulting in $B=2f_1 \lambda_0/3T_G$~\cite{dubost2018calibration}. Because of chromatic dispersion produced by the grating on diffraction mode $M_{+1}$, long and short wavelengths are pushed up and down respectively, when viewed as presented in Fig.~\ref{fig:layout}~(a). Consequently the shortest and longest wavelengths to fall inside the side aperture are $\lambda _\mathrm{min}=2\lambda_0/3=450\,\mathrm{nm}$ and $\lambda _\mathrm{max}=4\lambda_0/3=900\,\mathrm{nm}$, resulting in a maximum chromatic bandwidth of $66\%$ with respect to $\lambda_0$. A concern with shorter and longer wavelengths is that as they approach the edge of the side aperture, they are filtered asymmetrically by it. This will be discussed in Sec.~\ref{sec:poly}.

\begin{figure}[htb!]
  \centering\includegraphics[width=9.5cm]{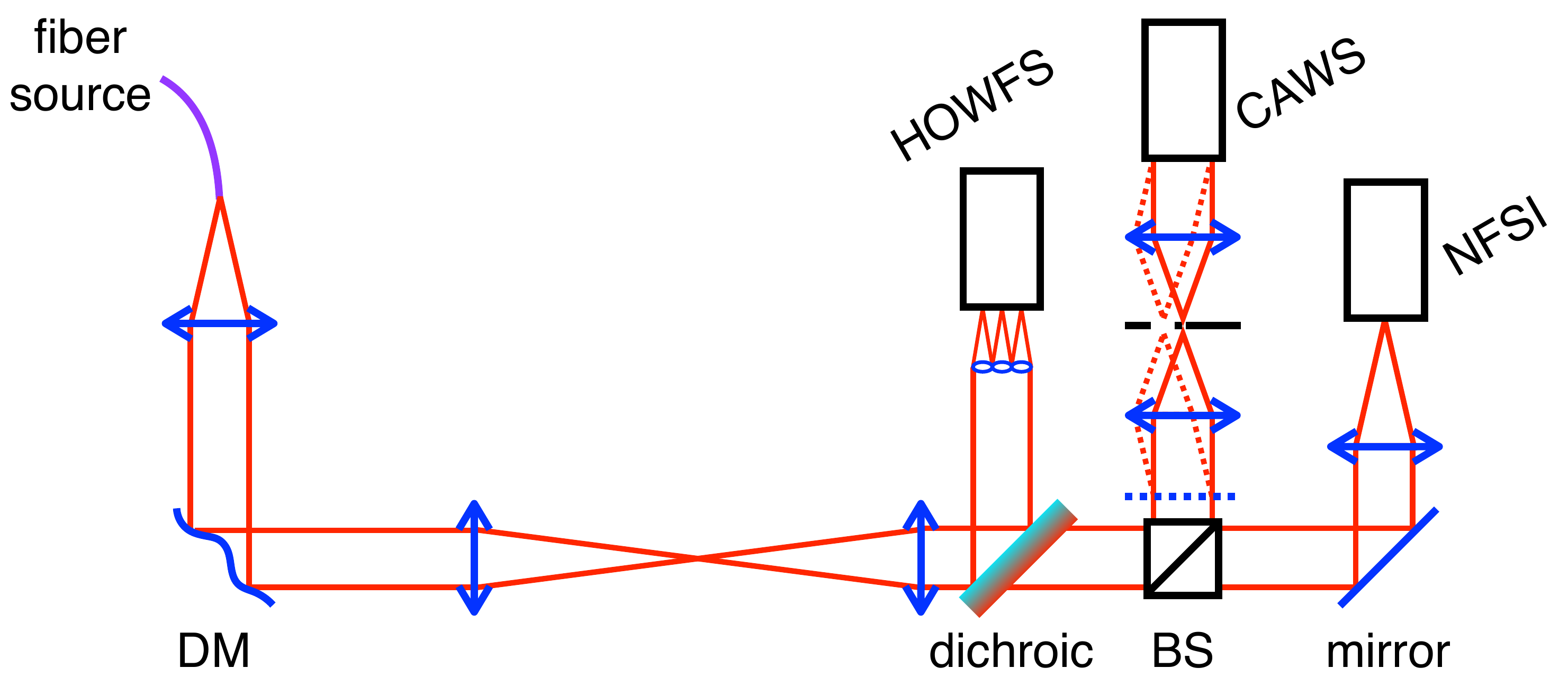}
  \caption{
  Schematic optical layout of the CHOUGH high-order AO system. The longpass dichroic is centered around $647\,\mathrm{nm}$. Both the HOWFS' lenslet array and the CAWS' grating are conjugated to the Deformable Mirror~(DM).}
  \label{fig:chough}
\end{figure}

The CAWS was hosted by the CHOUGH bench to produce the results presented in this paper. CHOUGH's layout is presented in Figure~\ref{fig:chough}.
The system comprises a 32-by-32 Boston Micromachines Kilo-DM which is used to both introduce and correct aberrations. Three instruments are located in closed-loop setup with the~DM: the Narrow-Field Science Imager (NFSI), the High-Order WFS (HOWFS) and the CAWS. The HOWFS is a 31-by-31 Shack-Hartmann reconfigured to have a subaperture sampling of $9.8$\,pixel/subaperture. The main purpose of this WFS is to calibrate the results when measuring the CAWS' accuracy. Both the HOWFS and the CAWS are conjugated to the~(DM). The NFSI is an imaging camera that operates between wavelengths $640<\lambda <880\,\mathrm{nm}$. The function of this imager is to provide absolute independent correction measurements during closed-loop correction tests. Due to a small leak on the $647\,\mathrm{nm}$ longpass dichroic, all instruments can be illuminated using a 633\,nm HeNe laser to feed the fibre. This is the main light source for monochromatic operations. Broadband illumination is provided by a halogen lamp.

\section{Dynamic range}
\label{sec:range}
Besides testing the instrument's accuracy, precision and completeness, this paper also seeks to validate our models regarding the interaction of the sideband with aperture $M_{+1}$. From our previous paper~\cite{dubost2018calibration} we know that the largest spatial frequency the instrument can sample is $\kappa_{max}=1/3T$ in $\mathrm{m^{-1}}$, and that for this to be possible the aperture's size $B=2a/3=2 \lambda_0 f_1 / 3T$, where $f_1$ is the focal length of lens $\ell_1$. But apart from defining the spatial sampling, our models suggests that the aperture's size adds 2 extra limitations.

The first limitation is on the m-PDI's dynamic range, or more precisely on the largest wavefront slope it can sense. For any local slope $s$ where the absolute value $|s|>B/2f_1= \lambda_0 /3T$, locally the wavefront has a tilt which makes the majority of the light propagate outside of aperture $M_{+1}$, leading to it not being sensed. This effect is similar to the one used in pyramid WFSs, where the local slope determines the proportion of light in each quadrant~\cite{ragazzoni1996pupil}. Aperture $M_0$ is less sensitive to local tilts due to diffraction. Coming back to aperture $M_{+1}$, for a slope to stay within the dynamic range, it must follow $|s|<\lambda_0 /3T$. This means that reducing the period $T$ of the grating not only increases the sampling, but it also increases the dynamic range. For the case of a spatially sinusoidal wavefront $\varphi_m (x,y)=A_m \cos(2\pi \kappa x)$, where $\kappa$ is the spatial frequency and $A_m$ is the amplitude measured in meters, the absolute value of the slope is at most $\max(|\partial \varphi_m / \partial x|)=2\pi \kappa A_m$. For the instrument to correctly sense such a wavefront, then
\begin{equation} \label{eq:dynamicrange}
    A_m < \frac{\lambda_0}{6\pi T \kappa}.
\end{equation}
This equation will be tested in Section~\ref{sec:transfer}.

The second limitation is on the sensitivity to high spatial frequencies. Coming back to a sinusoidal wavefront $\varphi_r (x,y)=A_r \cos(2\pi \kappa x)$, where this time $A_r=A_m (2\pi /\lambda_0)$ and is measured in radians, Taylor's expansion of the electric field $\Psi_0=e^{i \varphi_r}$ is 
\begin{equation}
    \Psi_0 = \sum^{\infty}_{j=0} \frac{(i A_r \cos(2\pi \kappa x))^j}{j!}.
\end{equation}
This expression gives us access to different harmonics. For $j=1$ we get the first harmonic $iA_r \cos(2\pi \kappa x)$, whereas for $j=2$ and by using the trigonometric identity $\cos^2(\theta)=(1+\cos(2\theta))/2$, we get access to the second harmonic $-A^2_r \cos(4\pi \kappa x)/4$. 

The ratio of amplitudes between the second and the first harmonics is $A_r/4$. This means that for larger values of $A_r$, the contribution of the second harmonic increases relative to that of the first one. This second harmonic is filtered by the $M_{+1}$ aperture for $\kappa>1/6T=\kappa_{max}/2$. In other words, for all frequencies above half the maximum frequency, the second harmonic gets filtered out, reducing the instrument's sensitivity to them. For example, with $A_r=1\, \mathrm{rad}$, the ratio between them is 0.25, which in turn translates to a drop of about 20\% in the amplitude of the observed wavefront. This drop was first observed on the theoretical transfer functions presented in our previous paper~\cite{dubost2018calibration}, where it remained unexplained. We seek to reproduce this phenomenon in the following section, hence validating our models.

\section{Spatial transfer function}
\label{sec:transfer}
The spatial transfer function is a representation of the instrument's accuracy regarding different spatial frequencies. In principle, its measurement is simple: a pure tone of known amplitude, i.e. a single spatial frequency, is produced, in this case by the DM, and then measured by the instrument. This spatial frequency, which has the shape of a sinusoid across the pupil, is held static for the duration of the detector's exposure. The transfer function for any given frequency is the ratio between the measured amplitude of that frequency and the amplitude of the input generated by the DM.

\begin{figure}[ht!]
\centering\includegraphics[width=9.14cm]{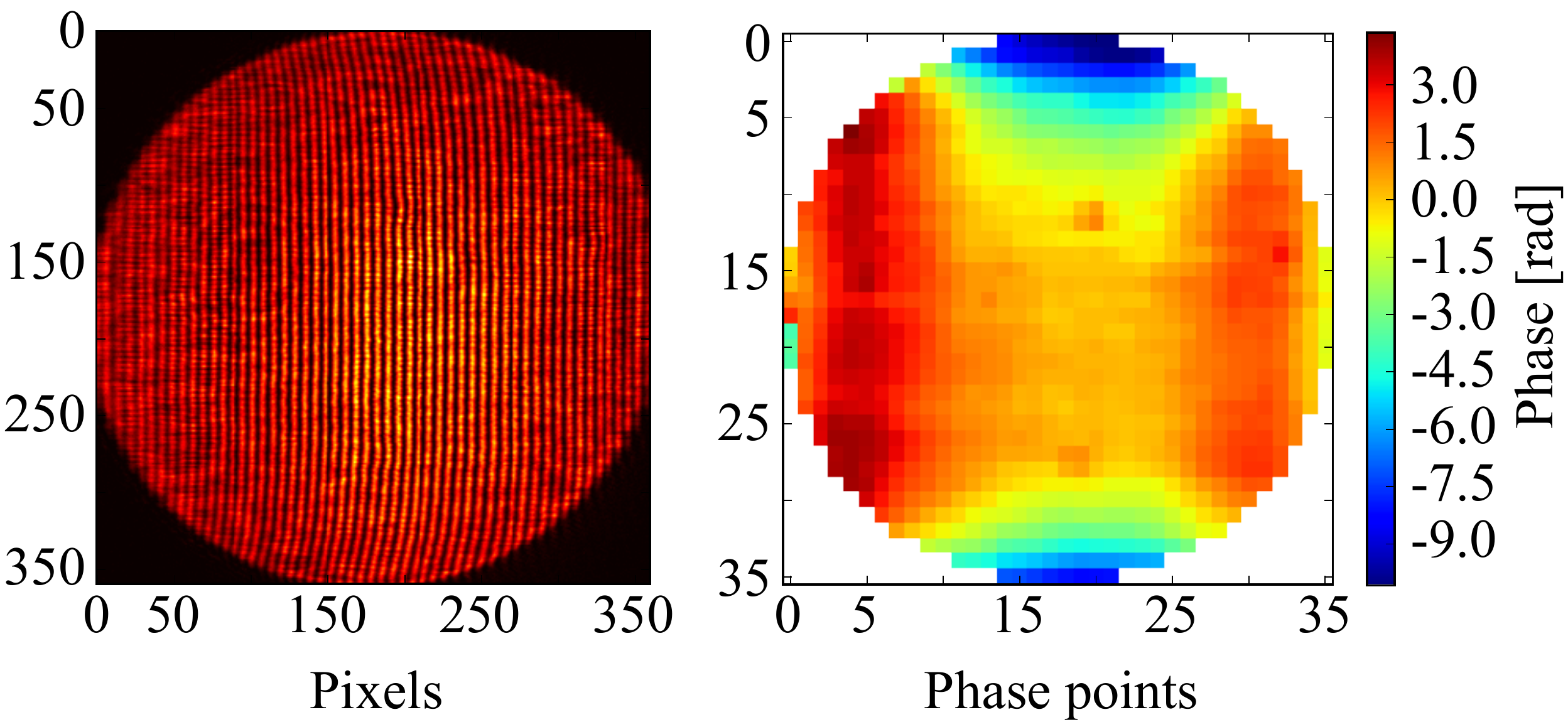}
\caption{(\textit{left}) Interference fringes produced with the DM in its zero position and (\textit{right}) the corresponding demodulated reference phase. The light source is a HeNe laser at $\lambda =633$\,nm. The physical scale of both images is the same as they span the length of the pupil, but the sampling is lower on the demodulated phase. The reason for this is that a line-pair is being sampled by approximately 7 pixels and the width of 3 line-pairs equals the width of 2 phase data points.}
\label{fig:choughReference}
\end{figure}

Before performing this experiment with CHOUGH, reference fringes are acquired to measure systematic biases and calibrate the data. Figure~\ref{fig:choughReference} shows monochromatic interference fringes and the corresponding demodulated phase. The data were obtained with the HeNe light source ($\lambda =633$\,nm) and with the DM sent to its zero position.
Note this position is a flat command on the DM at the middle of its range, which does not guarantee a flat shape. On the phase data, values go beyond the $\pm \pi/2$ and $\pm \pi$ limits which are characteristic to interferometers. This extension is achieved with a fast unwrapping algorithm~\cite{herraez2002fast}, which allows us to make full use of the CAWS' dynamic range. 2D single-frame phase unwrapping can be reliably achieved with a variety of algorithms for interferometers with a dynamic range of $\pm \pi$~\cite{ghiglia1998unwrap}, which gives us an advantage over other PDIs with ranges between $\pm \pi/2$. In this particular case, the peak-to-valley~(PtV) value of the measured phase is $5.1 \pi$\,rad, which is >2.5 times larger than the $\pm \pi$ limit and more than 5 times larger than for the $\pm \pi/2$ one. This shows that a good approach to increasing the small dynamic range of PDIs is to use a working principle that allows for a $\pm \pi$ range, such as the m-PDI, therefore facilitating single-frame phase unwrapping.

\begin{figure}[t!]
\centering\includegraphics[height=5.635cm]{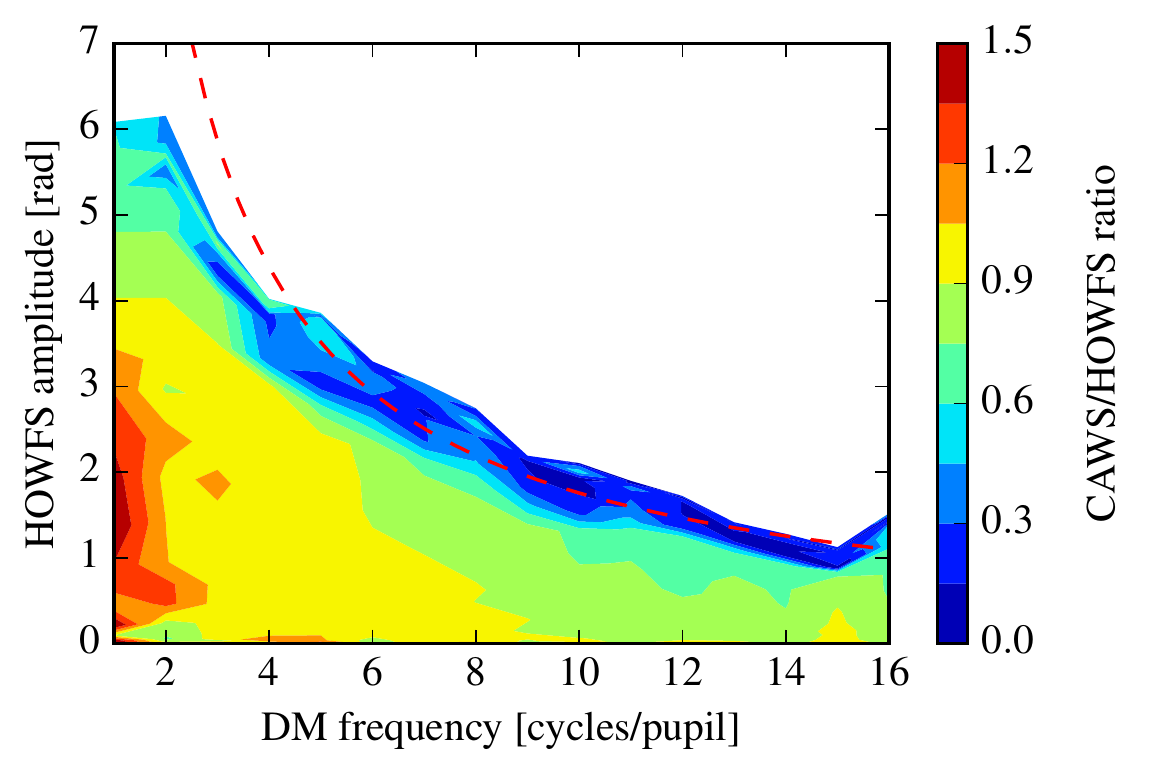}
\caption{CAWS' monochromatic transfer function, calibrated by the HOWFS for $\lambda =633$\,nm. The red dashed line marks the theoretical prediction of the CAWS' dynamic range as described by Eq.~\ref{eq:dynamicrange}.}
\label{fig:caws_howfs_ratio}
\end{figure}

After calibrating the biases, the CAWS is fed sinusoidal spatial aberrations in the X direction with the DM, spanning its full range of spatial frequencies, of up to 16\,cycles/pupil, and open-loop commanded amplitudes from 0 to $1.3\pi$\,rad~($2.6\pi$\,rad~PtV). The amplitudes actually achieved by the DM are determined by the HOWFS. In other words, the input amplitudes are calibrated with the HOWFS to compensate for the DM's non-linearities. Then, the transfer function is computed as the ratio between the CAWS's and the HOWFS' amplitude estimations, as shown by the contour plot in~Figure~\ref{fig:caws_howfs_ratio}.

The first thing to note on the figure is that there are no data points at high frequencies and high amplitudes, hence, the top-right half of the figure is empty. This happens because the DM can not access these amplitudes for those frequencies. Therefore, in the contour plot the frontier between the void and non-void part of the figure marks the maximum amplitude the DM can deliver. The decrease in maximum amplitude at higher frequencies is due to the inter-actuator coupling at high spatial frequencies, which prevents the DM from delivering these amplitudes. Above around 3 cycles/pupil, the maximum amplitude produced with the DM decreases proportional to the frequency.

It is possible to observe a tendency for the ratio to decrease from values above 0.9 and near 1 in the yellow area at the lower end of frequencies, down to values between 0.9 and 0.7 in the green  at the higher end. The total end-to-end drop of up to 30\% is in part due to a step drop taking place between 6 and 8~cycles/pupil, or about half way through the instrument's frequency range. This effect and the overall drop is in accordance to the loss of sensitivity predicted by the second diffraction order being filtered out, as presented in Section~\ref{sec:range}.

A small exception to this tendency seems to take place at a frequency of 15\,cycles/pupil, where the ratio increases back to above 0.9. From the raw data it is concluded that this small exception is due to local aliasing effects beginning to appear in both WFSs as they approach their spatial frequency limit. It is important to remember that local aliasing effects can occur at certain phases despite still satisfying Nyquist's sampling criterion. Since the CAWS can sample slightly higher frequencies than the HOWFS, it is less affected by aliasing at this particular frequency. In other words, this effect is produced by the HOWFS calibration of the data.

The last thing to note is that the ratios on the figure drop almost to zero alongside the red dashed line which predicts the CAWS' dynamic range, as described by Eq.~\ref{eq:dynamicrange}. The model matches the data for frequencies between 4 and 16\,cycles/pupil. Below these frequencies, the DM  does not provide sufficient amplitude to cover the dashed lined and test our model. In fact no physical DM can fully cover the dashed line at low frequencies, as doing so would require infinitely large amplitudes.
Regardless of these limitations, the DM had sufficient range and spatial resolution to allow the successful measuring of the instrument's dynamic range and accuracy through the construction of the spatial transfer function.

\section{Closed-loop}
\label{sec:loop}
\begin{figure}[ht!]
\centering\includegraphics[height=5.6cm]{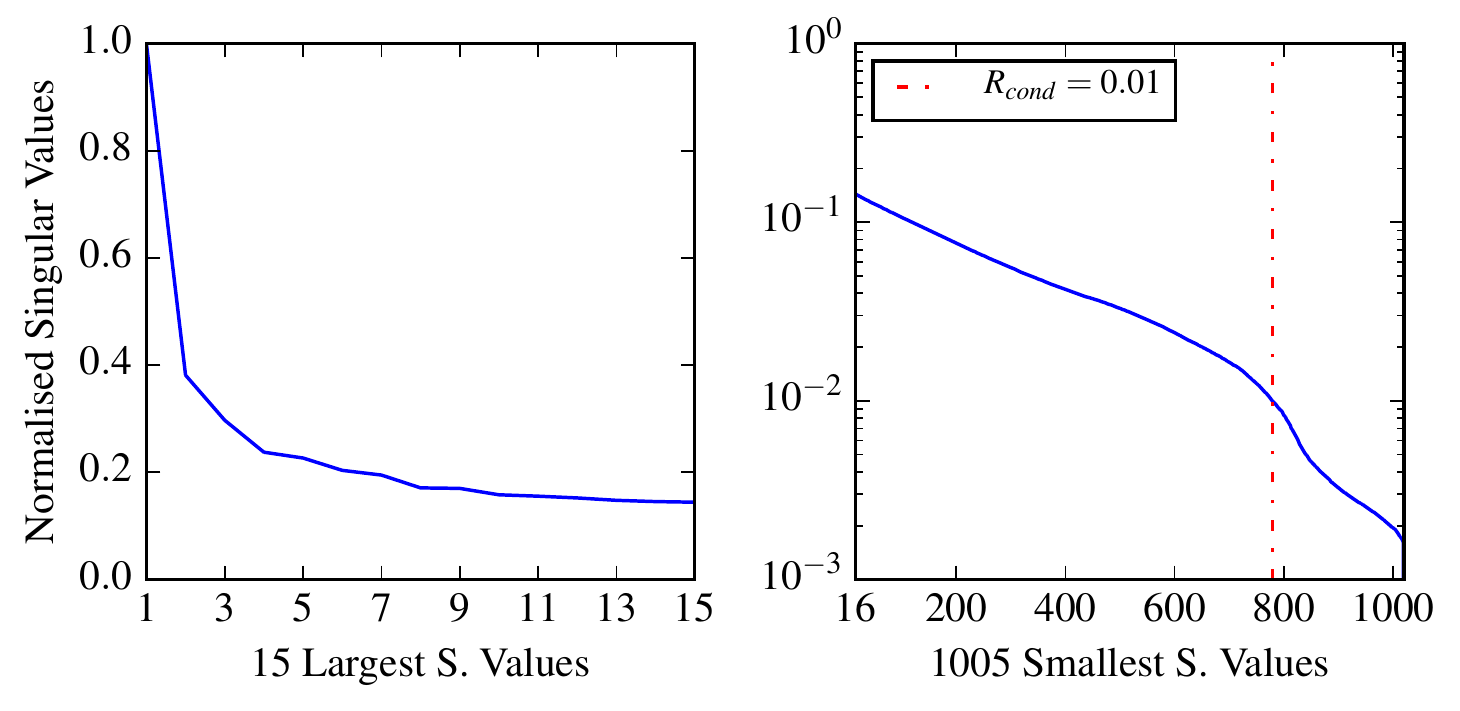}
\caption{Normalised singular values, sorted from highest to lowest. The \textit{left} panel shows the 15 largest values in detail and the \textit{right} panel shows the rest.}
\label{fig:singularValues}
\end{figure}

In order to assess the CAWS' completeness of measurements, the following experiment will test whether the CAWS can be used in closed-loop with CHOUGH's DM to correct for static aberrations. All tests in this section are performed monochromatically with the HeNe laser. For every control frame $k$, actuator commands $\overrightarrow{a_k}$ are computed using the control law
\begin{equation}
\overrightarrow{a_k} = \overrightarrow{a}_{k-1} + \alpha \, C_\mathrm{mat} \, \overrightarrow{s_k},
\end{equation}
where $\alpha$ is the loop's gain, $C_\mathrm{mat}$ is the control matrix, and $\overrightarrow{s}$ are WFS measurements. This matrix is built by poking all actuators simultaneously with an independent sinusoidal temporal frequency~\cite{kellerer2012deformable}, and then producing the pseudo-inverse of the resulting interaction matrix using singular value decomposition. Figure~\ref{fig:singularValues} shows the singular values obtained, sorted in decreasing order, and the threshold set by the conditioning.
As can be seen in the figure, 779 out of 1020~(total number of actuators: active and inactive) singular values are kept after thresholding. This threshold was iteratively found to produce the lowest residual error after closing the loop. With 801 active (illuminated) actuators, keeping 779 singular values means that we almost reached the maximum number of singular vectors that can contribute more useful information than they do noise. This is a sign that the CAWS measures signals from the DM with a high degree of fidelity.

\begin{figure}[ht!]
\centering\includegraphics[height=3.76cm]{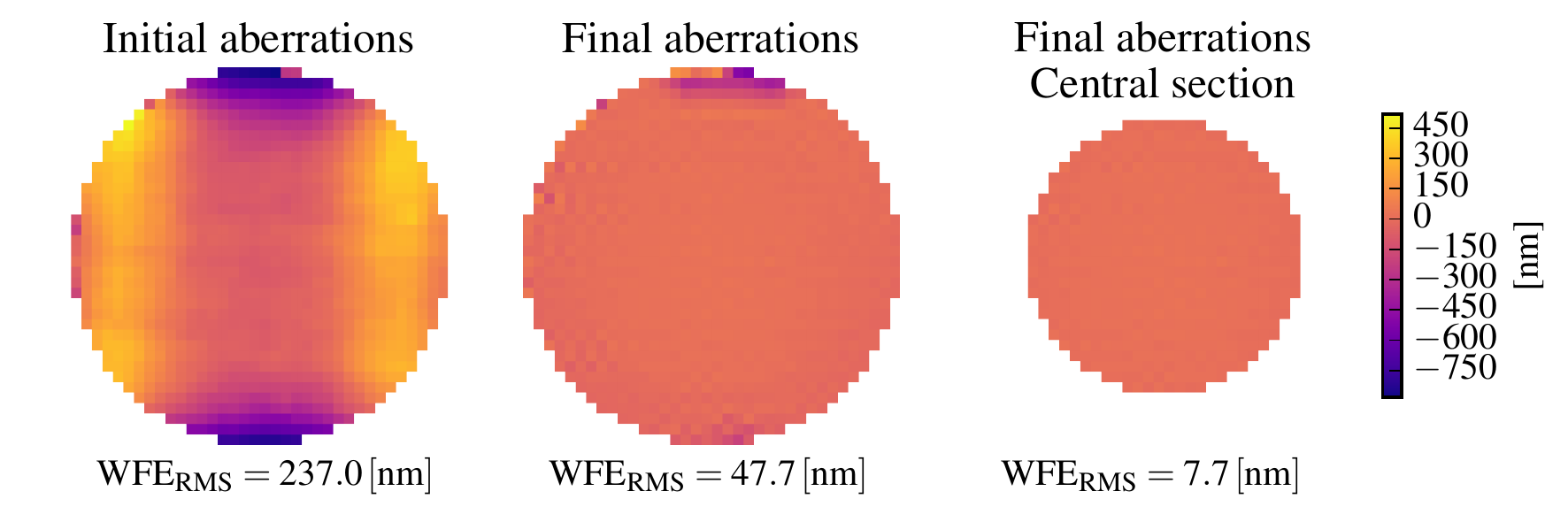}
\caption{CAWS phase measurements of (\textit{left}) the initial static aberrations, (\textit{centre}) the final aberrations the system converges to after closing the control loop, and (\textit{right}) a section at the centre of the pupil for those central aberrations.}
\label{fig:maskedPhase}
\end{figure}

Figure~\ref{fig:maskedPhase} shows the phase measured with the CAWS before and after closing the control loop on the systematic biases and waiting until it converges.
CHOUGH's initial static aberrations, as measured by the CAWS, have a RMS value of 237.0\,nm. If the entire pupil is considered, after closing the control loop the final RMS value of residual aberrations is 47.7\,nm, or 0.075 wavelength cycles. Note that most residual aberrations are located near the top and bottom edges of the pupil. In these regions the DM is fixed to its mount and so it cannot achieve the same range as with other actuators. This can be addressed by masking the reconstructed pupil to reduce its radius by $\sim 4.5$~actuators, or by 28\% of the total area as shown in the \textit{right} panel. Then the value of residual aberrations drops by more than a factor of 6 down to 7.7\,nm, or 0.012 cycles. This is equivalent to sensing and correcting an offset of~1.2\%~RMS in the position of the interference fringes or 0.08 pixels at a sampling of 6.9 pixels per line-pair, which is very encouraging for the first laboratory demonstration setup.

\begin{figure}[th!]
\centering\includegraphics[height=3.7cm]{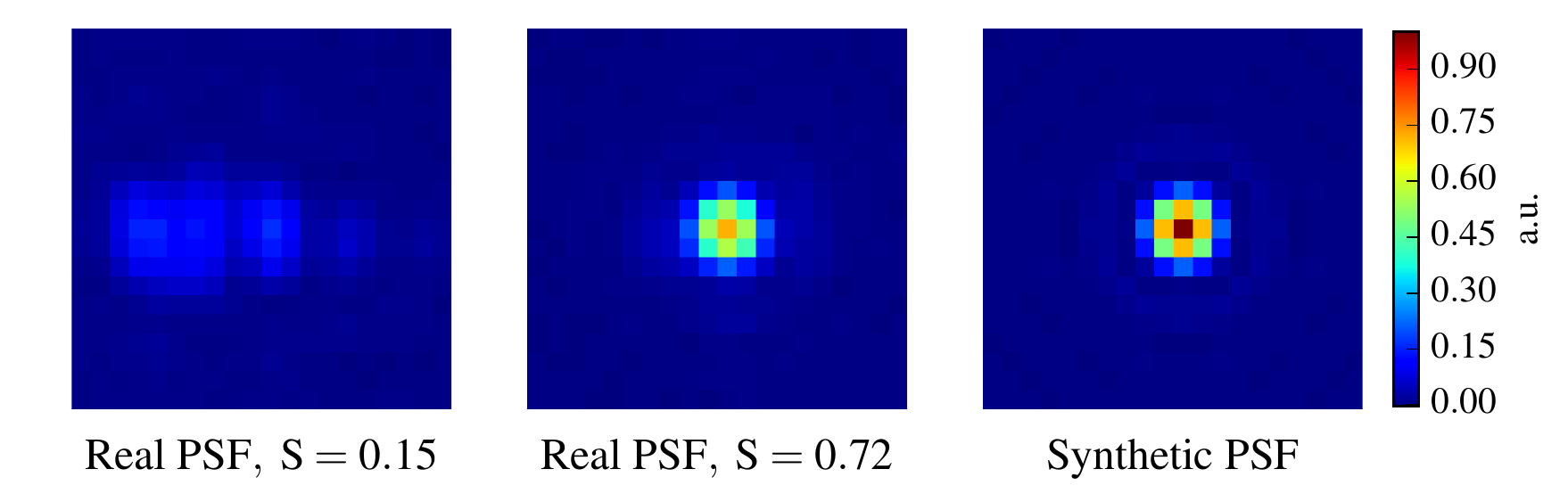}
\caption{NFSI focal plane images with (\textit{left}) the initial static aberrations, (\textit{centre}) the final aberrations the system converges to after closing the control loop, and (\textit{right}) the synthetic PSF used as reference to calculate SR, for comparison. All three images have been normalised to have the same cumulative intensity, and have been centred around a single pixel.}
\label{fig:psfImgs}
\end{figure}

Since the ultimate goal of AO is not to reduce the aberrations measured with a WFS, but rather to increase the angular resolution and brightness of focal plane images, then the best way to assess the performance of the control loop is by directly observing the value of metrics describing the quality of these images. This can be achieved by simultaneously acquiring images with CHOUGH's NFSI, and then estimating their Strehl ratio~(SR). Figure~\ref{fig:psfImgs} shows NFSI images before and after closing the control loop.
The central panel shows the last image acquired after letting the control loop converge and its corresponding SR. The average SR of the last 10 images is 0.73. The increase in~SR from the original 0.15 in the left panel of the figure confirms that the control loop is indeed removing sensed aberrations.

\begin{figure}[ht!]
\centering\includegraphics[height=5.49cm]{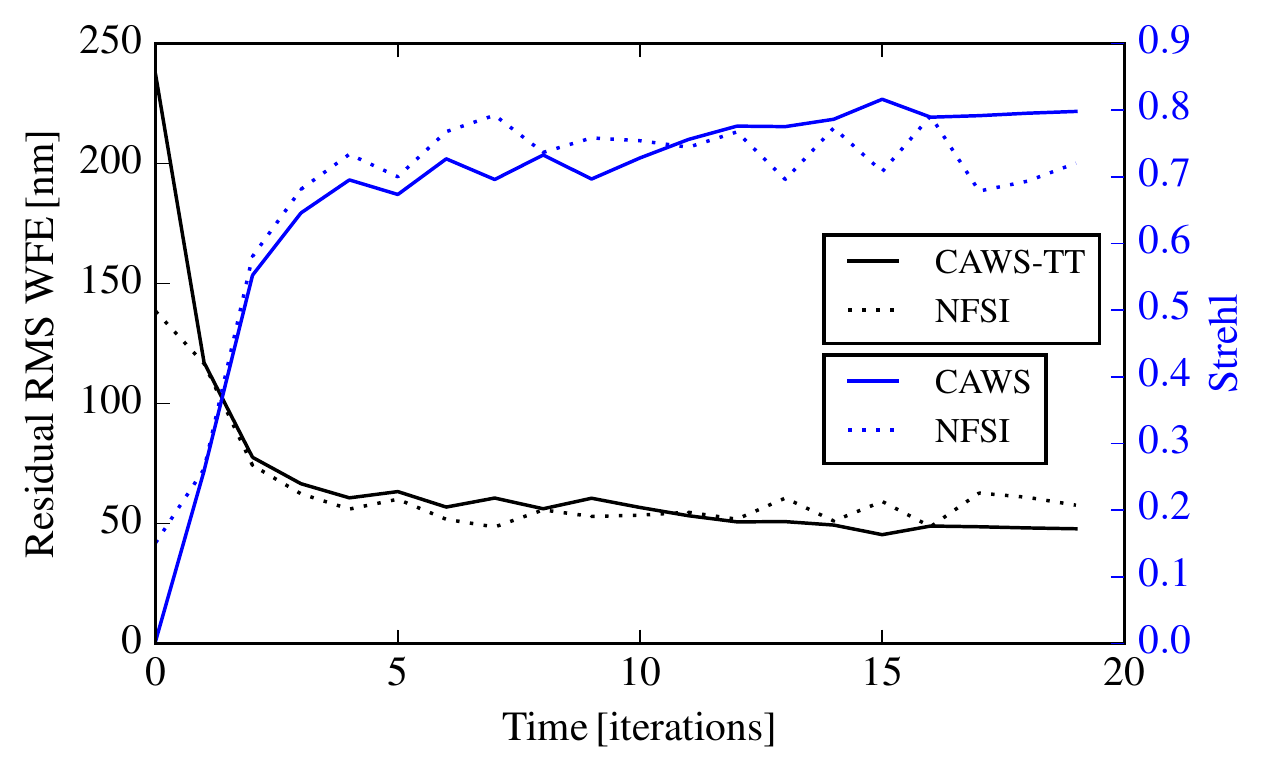}
\caption{Temporal evolution of residual RMS aberrations (\textit{black}) and SR (\textit{blue}) from the moment the control loop is closed until it converges. The SR was computed using NFSI images (\textit{dotted blue line}) and estimated using CAWS phase measurements (\textit{solid blue line}) minus tip-tilt and compensated for non-common path defocus. Residual tip-tilt was also subtracted from CAWS phase measurements before computing their RMS value (\textit{black solid line}). Finally, NFSI SR measurements and CAWS non-common path defocus estimations where combined using Eq.~\ref{eq:defocusncpa2}, to translate SR into RMS aberrations on the CAWS' path~(\textit{black dotted line}).
}
\label{fig:bestCurve}
\end{figure}

An advantage of the~SR is that it can be estimated from RMS aberrations through the use of Mar\'{e}chal's approximation~\cite{mahajan1982strehl}. This way, the aberrations on the NFSI's path can be compared to those on the CAWS. In order to estimate the SR on the NFSI using CAWS phase measurements, tip-tilt must be subtracted from them and non-common path defocus compensated for. Working under the assumption that the CAWS is difficult to focus before an observation, whereas the NFSI is easier to do so, then the NFSI's initial defocus can be considered to be negligible and any defocus seen by the CAWS before closing the loop would be non-common to both instruments. Hence, Mar\'{e}chal's approximation can be extended so that the SR can be estimated by the CAWS for any frame $i$ as
\begin{equation}\label{eq:defocusncpa}
    S_i=e^{-(\sigma_i^2 + (d_i-d_0)^2)(\frac{\lambda_c}{\lambda_n})^2},
\end{equation}
where $\sigma_i$ is the RMS value of all aberrations seen by the CAWS for frame $i$, minus tip-tilt and defocus, $d_i$ and $d_0$ are the amplitudes of the defocus term seen by the CAWS at frames $i$ and before closing the loop respectively, and $\lambda_c$ and $\lambda_n$ are the central wavelengths on the CAWS and the NFSI respectively. Note that in this particular case $\lambda_n$ equals $\lambda_c$. Conversely, Eq.~\ref{eq:defocusncpa} can be rearranged in order to estimate the RMS value of aberrations on the CAWS using a combination between NFSI SR measurements and CAWS defocus terms, so that
\begin{equation}\label{eq:defocusncpa2}
    \sqrt{\sigma_i^2 + d_i^2}=\sqrt{-\frac{\lambda_n^2}{\lambda_c^2}log(S_i)+2d_0d_i - d_0^2}.
\end{equation}
Figure~\ref{fig:bestCurve} shows the temporal evolution of both the RMS aberrations and the~SR after closing the loop, as estimated using both the CAWS and the NFSI.

As is shown in the figure, independent estimations of the RMS aberrations and of the~SR show good agreement after the first iteration of the control loop. This is strong proof of the completeness of the CAWS as a WFS, showing that there are no significant aberrations that are not seen by it. With respect to the initial disagreement, this is due to the system being in a low~SR regime~(0.15) where, as shown by~\cite{mahajan1983strehl}, Mar\'{e}chal's approximation underestimates the RMS value of aberrations.

\section{Broadband spectrum}
\label{sec:poly}
After successfully tuning and closing the loop with monochromatic light, the same task was carried out with polychromatic broadband light produced with a halogen lamp. In combination with the 647\,nm dichroic, the resulting spectrum was centered around $\lambda_\mathrm{C}=687\,\mathrm{nm}$ and had a FWHM of $\Delta \lambda_\mathrm{FWHM}=77\,\mathrm{nm}$, or about 18\% of the total theoretical bandwidth of 450\,nm. This is about half the bandwidth of the R~band and is centered around a comparable wavelength.

A concern was all but the central wavelength would be asymmetrically filtered by $M_{+1}$, the consequences of which have not yet been theoretically modelled. By limiting the bandwidth to 18\% we expect to limit the effects of this asymmetrical filtering.

Another chromatic consideration is that for a wavelength independent OPD, the wavefront's phase can be rewritten as $\varphi_{LP} = OPD \cdot \lambda / 2 \pi$. When replaced into Eq.~\ref{eq:simpleic}, this makes the modulation of interference fringes a function of the wavelength. Fortunately and as modelled in our previous paper~\cite{dubost2018calibration}, this should not have an effect in the estimated phase when all wavelengths have the same intensity.

\begin{figure}[ht!]
\centering\includegraphics[height=5.49cm]{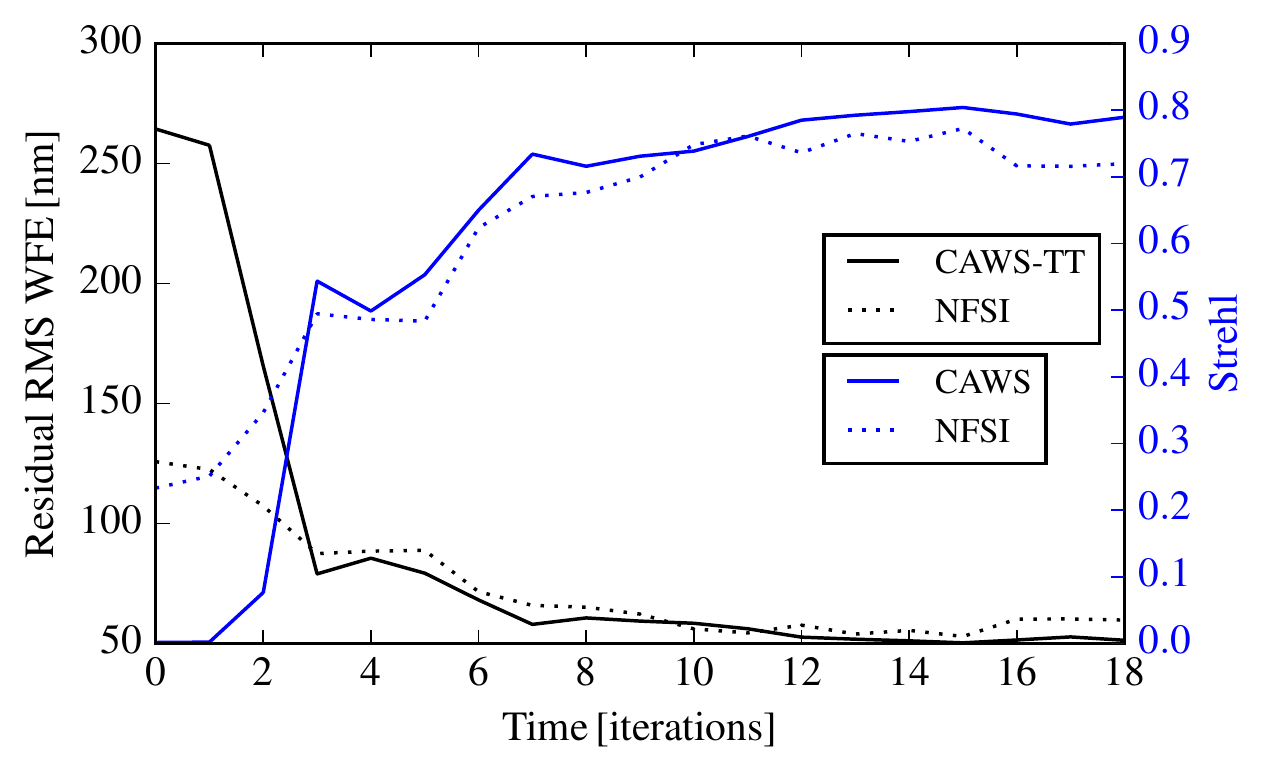}
\caption{Temporal evolution of (\textit{black}) residual RMS aberrations and (\textit{blue}) SR from the moment the control loop is closed until it converges. These results correspond to the same experiment presented in Figure~\ref{fig:bestCurve}, but with a broadband halogen lamp rather than a HeNe laser. The equivalent central wavelength is 687\,nm and the bandwidth is 77\,nm.}
\label{fig:bestPolyCurve}
\end{figure}

To measure the SR, the NFSI was fitted with a narrow band filter centred around 655\,nm and with a bandwidth of 15\,nm. Results are presented in Figure~\ref{fig:bestPolyCurve}. As shown in the figure, the final values for the~SR and the RMS~residual aberrations of 0.74 and 51\,nm respectively are almost identical as they were in the monochromatic case. The last 10 frame average SR is also 0.74. This shows that the CAWS can accurately and precisely sense aberrations with at least this chromatic bandwidth.

\section{Discussion}
\label{sec:diss}
Besides introducing the first experimental demonstration of the m-PDI, this paper presents the results of three experiments aimed at empirically validating this concept. The characterisation of the transfer function determines the CAWS' accuracy, the closed-loop experiment tests its precision and completeness and the polychromatic measurements confirm that the instrument can perform accurate and precise broadband light wavefront sensing.

With respect to the characterisation of the transfer function, our measurements confirm both our original hypotheses regarding the dynamic range being limited by the sideband aperture size, and the drop in sensitivity at high spatial frequencies due to the filtering of the second harmonic. The most important consequence of these phenomena is that both the dynamic range and the sensitivity at high frequencies can be extended by increasing the size of the sideband aperture $M_{+1}$. This would of course also increase the instrument's spatial sampling, which would have to be accompanied by an increase in the number of pixels across the pupil by the same factor.

In the closed-loop tests, the final residual aberrations seen by the CAWS across the entire pupil and inside a central section were 47.7\,nm~rms and 7.7\,nm~rms respectively. This result is significant, as it places the CAWS alongside other PDIs capable of achieving nanometric accuracy, but with an extended single-frame dynamic range and chromatic bandwidth.

Finally, the ultimate goal of validating the CAWS' chromatic bandwidth is to show it can perform wavefront sensing on a natural guide star without needing narrowband filtering (of a few nanometers wide) which would restrict the instrument's overall throughput. The next step is then to attempt such measurements at the back of an AO system with XAO capabilities and ample room for visiting instruments.

\section{Acknowledgment}
We thank Nigel Glover, the Department of Physics, and the University of Durham for granting us safe laboratory access during the Covid-19 pandemic, in order to complete all the measurements for this publication.

\section*{Funding}
Comisi\'{o}n Nacional de Investigaci\'{o}n Cient\'{i}fica y Tecnol\'{o}gica (CONICYT) (72160371); UK~Research and Innovation Science and Technology Facilities Council (STFC) (ST/L002213/1,  ST/L00075X/1, ST/P000541/1, ST/T000244/1).

\section*{Disclosures}
The authors declare no conflicts of interest.

\section*{Data availability}
Data underlying the results presented in this paper are not publicly available at this time but may be obtained from the authors upon reasonable request.

\bibliographystyle{IEEEtran}
\bibliography{scholar}
\end{document}